\documentclass[aps,amsmath,amssymb,nofootinbib,preprintnumbers]{revtex4}
\usepackage{graphicx}
\usepackage{epstopdf}
\usepackage{amsmath}
\usepackage{amsfonts}
\usepackage{amssymb}
\usepackage{enumerate}
\usepackage{amsthm}
\usepackage{array}
\usepackage{epsfig}
\usepackage[small,bf]{caption}
\usepackage{appendix}
\usepackage{dsfont}
\usepackage{hyperref}
\def\1ad{\mbox{\normalsize $^1$}}
\def\2ad{\mbox{\normalsize $^2$}}
\def\3ad{\mbox{\normalsize $^3$}}
\def\4ad{\mbox{\normalsize $^4$}}
\def\5ad{\mbox{\normalsize $^5$}}
\def\6ad{\mbox{\normalsize $^6$}}
\def\7ad{\mbox{\normalsize $^7$}}
\def\8ad{\mbox{\normalsize $^8$}}

\def\beq{\begin{equation}}                     % 
\def\eeq{\end{equation}}                       %
\def\bea{\begin{eqnarray}}                     %         %
\def\eea{\end{eqnarray}}                       %       % 
\def\dj{\hbox{d\kern-0.347em \vrule width 0.3em height 1.252ex depth
-1.21ex \kern 0.051em}}

\def\ket{\rangle}
\def\bra{\langle}

\def\Dirac{\,\raise.15ex\hbox{/}\mkern-13.5mu D}
\def\dirac{\,\raise.15ex\hbox{/}\kern-.57em \partial}
\def\pslash{\,\raise.15ex\hbox{/}\kern-.57em p}

\def\CO{{\cal O}}
\def\ellp{\ell_{\rm P}}

\begin{document}

\begin{flushright}
IFT UAM/CSIC-11-28
\end{flushright}

\title{ Chaotic Fast Scrambling At Black Holes}

\author{{\sc Jos\'e L.~F.~Barb\'on}\footnote{jose.barbon@uam.es} and {\sc Javier M.~Mag\'an}\footnote{javier.martinez@uam.es}}
\affiliation{Instituto de F\'{\i}sica Te\'orica  IFT UAM/CSIC,
 Campus de Cantoblanco, E-28049
Madrid, Spain \\}

\date{May 2011}

%Abstract
\begin{abstract}

Fast scramblers  process information in characteristic times scaling logarithmically with the entropy, a behavior which has been conjectured for black hole horizons. In this note we  use the AdS/CFT fold to argue that  causality bounds on information flow  only depend on the properties of  a single thermal cell, and admit a geometrical interpretation in terms of the optical depth, i.e. the thickness of the Rindler region in the so-called optical metric. The spatial sections of the optical metric are well approximated by constant-curvature hyperboloids. We use this fact to  propose an effective kinetic model of scrambling which can be assimilated to a compact hyperbolic billiard, furnishing a classic example of hard chaos.  It is suggested that classical chaos at large $N$ is a crucial
 ingredient in reconciling the notion of fast scrambling with  the required saturation of causality.

\end{abstract}

\pacs{11.25.Tq, 05.30.Ch}

\maketitle

%%%%%%%%%%%%%%%%%%%%%%%%%%%%%%%%%%%%%%%%%%%%%%%%%%%%%%%%%%%%%%%%%%%%%%%%%%%%%%%%%%%%%%%%%%%%%%%%%%%%%%%%

\setcounter{equation}{0}

\section{Introduction}
\noindent

It was proposed in \cite{Hayden, Sekino}  that one defining characteristic of black holes as quantum systems is their very fast rate of  information scrambling. In systems with local degrees of freedom and local  interactions   the scrambling of information can often be  assimilated to   a diffusion process. In this case, we can say that locally coded information  `random-walks' around the system, covering a volume $L^d$ in a time proportional to $L^{2}$.  For `normal' states, with extensive entropy, $S\propto L^d$, we find scrambling times proportional to  $S^{2/d}$, measured in units of some typical time scale characterizing the state, such as the inverse temperature $\beta=T^{-1}$ in relativistic systems.  In contrast, the scrambling time for black holes is conjectured to be 
$$
\tau_s \sim T^{-1} \,\log(S)\;,
$$
 with $T$ the Hawking temperature and $S$ the black hole entropy. Roughly, it corresponds to the
local scrambling for a system with infinite spatial dimension, $d\rightarrow \infty$. It is very interesting to elevate this estimate to the rank of 
fundamental law of nature, and regard black holes as the fastest possible scramblers \cite{Sekino}. 

The fast scrambling rate of black holes can be argued in a number of ways. It turns out that $\tau_s$ is the time scale
that saturates the no-cloning bound in black holes, i.e. it is a measure of the minimum time for information retrieval in the Hawking radiation (cf. \cite{Hayden, Sekino}). The same time scale can be identified in a kinematical effect characteristic of the `membrane paradigm'  \cite{Damour, Thorne, Susskindbook}, where a conserved  charge distribution on the stretched horizon, induced by the motion of external charges, undergoes non-relativistic Ohmic diffusion with respect to the Schwarzschild time variable. It follows that this induced charge spreads exponentially fast, filling a given area of horizon in a time which scales logarithmically  with this area \cite{Sekino}.
Finally, from the point of view of hypothetical holographic duals of the black hole, both Matrix Theory \cite{Matrix} and AdS/CFT \cite{adscft}  
suggest the consideration of  $N\times N$ matrix models with $S\sim N^2$ degrees of freedom. The non-local character of the interactions of matrix elements in `index space' suggests that,  precisely for these systems,  we may get the maximal possible scrambling rate.

It is somewhat  striking that the two heuristic determinations of the $\tau_s$ time scale, based respectively on the no-cloning argument and the membrane paradigm, should look so different at  a superficial level. In this note, we start by reinterpreting these arguments as causality bounds in a theory with a stretched horizon. 
In particular, it is  pointed out that the no-cloning time is just the minimal reflection time from  the stretched horizon, across the
near-horizon region. With this definition, we analyze a completely general near-horizon system embedded in a generalized AdS/CFT background and find a universal answer for such a  `reflection' time:
\beq\label{univ}
\tau_*= \beta\,\log\,(S_{\rm cell})\;,
\eeq
where  $\beta = T^{-1}$ and  $S_{\rm cell}$ is the entropy contained in a single `thermal cell' of the dual CFT, a region with the  size of  the thermal length, $\beta=T^{-1}$, measured in the CFT metric.

On general grounds, the idea that information should scramble `as fast as the speed of light' seems counterintuitive, motivating the view that no local or quasilocal model of the stretched horizon can describe such a phenomenon. 

In this note we propose a refinement in this conceptual tension. 
We argue that a purely kinetic phenomenological model can behave, in a certain sense,  as a fast scrambler provided the scrambling is achieved in a few collisions, as in a classically chaotic system.  The route to classical chaos in the near-horizon geometry is actually found in the conformally-related optical metric, which approximates any near-horizon region as a constant-curvature hyperboloid. 

While Rindler space is metrically  equivalent to  flat Minkowski space, there are many dynamical properties, which depend on {\it phase space} rather than configuration space, which are better characterized in terms of the optical metric. In particular, the non-compactness of the optical metric is directly related to the non-compactness of the phase space at black hole horizons. The stretching of the horizon can be seen as a compactification of phase space which renders black-hole entropy finite. Our analysis shows that these ideas tile beautifully with the chaotic nature of scrambling, since a compact hyperboloid is a classic paradigm of hard chaos.  \footnote{See \cite{pando} for another analysis of chaos in AdS/CFT, from quite a different point of view.}

\section{ Optical Depth And Causality Bounds}
    
\noindent

In this section we interpret the time scale $R_s \log(S)$ in the near-horizon physics of an ordinary Schwarzschild black hole in terms of a causality bound. 
We begin with the no-cloning argument (see  \cite{Hayden, Sekino}) by noticing Figure 1, which illustrates the fact that the edge of the near-horizon region, given by the line $X^+ X^- =-R_s^2$, is symmetric to the singularity locus, $X^+ X^- = R_s^2$, by a null reflection through the horizon. The points $Q$ and $Q'$ are also symmetric and separated a proper distance of order $\ellp$ from the horizon.

If a Q-bit falls into the black hole from $P$ to $Q$ while it is cloned at the horizon and returns within the Hawking radiation at $P'$,  a local verification of the cloning could occur inside the black hole if $Q$ and $P'$ can have a common point $S$ in their future. Notice that the points $P$ and $P'$ are set at height $R_s$ because the cloned Q-bit must be efficiently detected while it is emitted with wavelength of $\CO(R_s)$. This situation being incompatible with the linear evolution of quantum mechanical states, it must be that such a local meeting of the two copies is prevented by the occurrence of the singularity. In this way we obtain   lower bound on the time delay between the two departures of the infalling Q-bit from  $P$ and $P'$.

\begin{figure}
  \label{clones}
  \begin{center}
    \epsfig{file=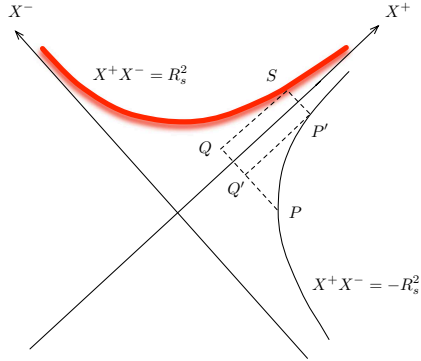, width= 5.5cm}
    \caption{ \small Diagram showing that  the Schwarzschild time between $P$ and $P'$ equals  the reflection time of a photon from the stretched horizon, i.e. the piecewise-null trajectory $P\,Q' \cup Q' \,P'$, where $Q$ and $Q'$ lie respectively at the inner and outer edges of the stretched horizon, i.e.  on the hypersurfaces $X^+ X^- = \pm \ellp^2$. }
  \end{center}
\end{figure}

We compute this delay using  the standard map between Schwarzschild time and Kruskal coordinates,  $X^\pm \sim R_s \exp(\pm t/ 2R_s)$. Since $Q'$ sits at the edge of the stretched horizon we have $X^+_{Q'} X^-_{Q'} = -\ellp^2$. Furthermore, $X^+_{P'} X^-_{P'} = -R_s^2$ and, by the reflection conditions $X^-_S =X^-_Q =- X^-_{P'} = -X^-_{Q'} $, we find
$$
\Delta t_{\rm min} \sim 2R_s \,\log \left({X^+_{P'}\over X^+_P}\right)  =2R_s\, \log \left({R_s^2 \over \ellp^2}\right).
$$
Noting that the entropy $S\sim (R_s / \ellp)^2$ we find, up to $\CO(1)$
coefficients and neglecting $\CO(R_s)$ additive contributions, 
$$
\Delta t_{\rm min} \sim \tau_* =  R_s\,\log(S)\;.
$$
The time-delay scale, $\tau_*$, is equal to the  `reflection time' from the stretched horizon or, in order of magnitude, a minimal free-fall time to the stretched horizon across the Rindler region. We shall denote this kinematical time scale as the `optical depth'.\footnote{The coincidence between the conjectured scrambling time and the free-fall time was recently noticed in \cite{Susskindnew}.}
Since $\ellp$ enters logarithmically,  this time scale is very robust in order of magnitude. For example, the same
result  is obtained with the much  less conservative criterion that $X^-_Q > M^{-1}$, with $M$ the mass of the black hole, so that the infalling Q-bit is not heavier than the original black hole.

The actual relevance of these arguments for the issue of information retrieval is not  clear.  First, it must be noted that the  `same bit' can be extracted from a few Hawking quanta provided one knows the black hole microscopic state with precision \cite{Hayden}, something that requires monitoring the black hole for a time of the order of the evaporation time (cf. \cite{page}).  Even then, the complete {\it decoding} of the information contained in a single Q-bit is likely to require time scales in excess of  $\CO(R_s \,e^S)$,  the Heisenberg time of the system 
controlling the detailed recurrences of the full quantum state (cf. \cite{maldar, susr, recc, russ, porrati, vijay}). It is perhaps more appropriate  to interpret  $\Delta t_{\rm min}$  as the {\it causality} bound on the  minimum time for the information to be returned to the edge of the Rindler region, i.e. the minimum time for the information to `resurface' in the outer layers of the black-hole's thermal atmosphere.  Notice that, while the causality bound is saturated by a `mirror' boundary condition at the stretched horizon,  the no-cloning heuristic argument allows the `processing' of the Q-bits at the stretched horizon to introduce further delays of $\CO(R_s)$,  or even of  $\CO(\tau_*)$, without affecting the  bound in order of magnitude.

A second argument introducing the time scale $\tau_* = R_s\,\log(S)$ uses the dynamics of induced charge densities on the stretched horizon according to the membrane paradigm  \cite{Damour, Thorne, Susskindbook, Sekino}. These charges are nothing but rescaled versions of the normal electric field at the stretched horizon, and their spacetime variations cannot violate local causality. The fastest possible variations of such electromagnetic fields can be induced by letting a moving source charge above the horizon approach the speed of light, leading naturally to the time scale $\tau_*$, since that is the characteristic time for the source charge to cross the Rindler region.  To see that an exponentially fast spread,  as measured  in asymptotic time, is compatible with causality we may recall a simple argument from \cite{susskindholo} showing that such a behavior is precisely {\it saturating } the causality bound, rather than contradicting it. Let us define the stretched horizon in locally flat coordinates  $(X^+, X^-, X_\perp)$ as the Rindler surface 
$$
X^+ X^-  \equiv -(X^0)^2 + Z^2  =  \ellp^2\;.
$$
The future light-cone bounding causal events originating at a point of coordinates $(X^0=0, X_\perp =0, Z=\ellp)$ on  the stretched horizon is given by
$$
-(X^0)^2 + (Z-\ellp)^2 + X_\perp^2 =0\;,
$$
which intersects the stretched horizon itself along the hypersurface 
$$
X_\perp^2 = 2\ellp(Z-\ellp) \longrightarrow 2\ellp^2\,e^{t/2R_s}
\;,
$$
where we have inserted the long-time asymptotics of $Z\sim X^+ - X^-$, with respect to the asymptotic Schwarzschild time variable, showing the exponential rate which fills an area of order $X_\perp^2 \sim R_s^2$ in a time of order $\tau_*$. Notice that the light rays between two points on the stretched horizon do not propagate confined to the stretched horizon, but actually go {\it above} it through the bulk of the Rindler region, a simple fact that will be crucial in what follows.

These arguments leave us in a quandary regarding the interpretation of $\tau_*$ as a scrambling time. Since
any process measured by scales of order $\tau_*$ is saturating causality in order of magnitude, this means that the purported scrambling must be not only fast but in fact as fast as it can be. Hence the suggestion that fast scramblers cannot behave like standard local diffusion at all: a fast scrambler must work at {\it ballistic} speed. Meanwhile, interactions of probe Q-bits with the Hawking radiation are small at least for a time of order $\tau_*$, since the `fall' is well approximated as a {\it free fall}.

\subsection{Optical Depth And The Thermal Cell}

\noindent

In this section we give an estimate of the optical depth for a general near-horizon geometry with a putative holographic dual. The main purpose of this exercise is to express the result in terms of dual QFT variables in a situation where  the AdS/CFT lore may be applied.

Consider a general $(d+2)$-dimensional background which may admit a holographic interpretation in terms of a $(d+1)$-dimensional  field theory at finite temperature:
\beq\label{holob}
ds^2 = F(\rho)\left(-h(\rho) dt^2 + d\ell^2 \,\right) + {d\rho^2 \over h(\rho)}\;,
\eeq
where  $\rho$ is the holographic radial coordinate, parametrizing the UV regime at large $\rho$ and the IR regime at small $\rho$. If the model has the luxury of being defined as a perturbation of a well-defined UV fixed point, the warp factor has the asymptotic behavior $F(\rho) \rightarrow \exp(2\rho/R_\infty)$ as $\rho\rightarrow \infty$, with $R_\infty$ the asymptotic AdS radius of curvature. The metric of the dual QFT is defined as the asymptotic induced metric after removing the $F(\rho)$ warp factor, i.e.
 \beq\label{qftm}
 ds^2_{\rm QFT} = -dt^2 + d\ell^2
\;,\eeq
where $\ell$ parametrizes the $d$-dimensional spatial manifold on which the QFT is defined. 
The function  $h(\rho)$  models   thermal effects, having a simple zero at the horizon, $h(\rho_0)=0$,  and
approaching unity at large values of $\rho$, far from the horizon.
The near-horizon, or Rindler region, reaches out to radii of order $\rho_\beta$, defined by $h(\rho_\beta) \sim 1$. In the linear approximation we have $h(\rho_\beta) \sim h'_0 (\rho_\beta- \rho_0) \sim 1$, which gives an order-of-magnitude estimate for $\rho_\beta$.  The Hawking temperature is given by
$$
T= {h'_0 \over 4\pi} \sqrt{F_0}\;,
$$
where $F_0 \equiv F(\rho_0)$. 

Throughout our discussion, we assume that the temperature is much larger than any other energy scale in the dual QFT state, so that we can neglect finite-size effects in the geometrical set up. The stretched horizon is set at Planck distance from the horizon, at coordinate $\rho_*$ defined by
$$
\ell_{\rm P} = \int_{\rho_0}^{\rho_*} {d\rho \over \sqrt{h(\rho)}} \sim \sqrt{\rho_* - \rho_0 \over h'_0}\;.
$$
Using $h_* = h(\rho_*) \approx h'_0 (\rho_* - \rho_0)$ we find the relation 
$$
\beta \sqrt{F_0 h_*} \sim \ell_{\rm P}\;,
$$
 which expresses the fact that the local blue-shifted temperature, proportional to $(F(z)h(z))^{-1/2}$,  grows from $\CO(T)$ at the edge of the Rindler region, to Planckian order $\CO(m_{\rm P})$ at the stretched horizon. 

Null radial trajectories are better described in terms of the Regge--Wheeler coordinate, $z$ defined by
$$
dz = -{d\rho \over h(\rho)\sqrt{F(\rho)}} \;,
$$
where the minus sign is meant to indicate that $z$ grows  `inwards'. The metric in this  frame reads 
\beq\label{rwhe}
ds^2 = F(z) h(z) \left(-dt^2 + dz^2 \right) + F(z) \,d\ell^2 \;, 
\eeq
so that null radial trajectories satisfy $dz/dt=\pm 1$, and the optical depth can be measured directly by the extent of the $z$ coordinate. The same null trajectories occur in the conformally-related metric 
\beq\label{opme}
ds_{\rm op}^2 = -dt^2 + dz^2 + h(z)^{-1} \,d\ell^2 
\;,\eeq
 so-called `optical metric' \cite{optical}. With an appropriate choice of the additive normalization for the $z$ coordinate, we find that
 the optical metric of the holographic background (\ref{holob}) has the following {\it universal} structure: the asymptotic vacuum region is represented by a flat strip of length $\beta$, 
 \beq\label{asymop}
 ds_{\rm op}^2 \approx -dt^2 + dz^2 + d\ell^2\;, \qquad 0<z<z_\beta \sim \beta\;, \eeq
 with optical depth of order $z_\beta \sim \beta$. The near-horizon or Rindler region is well-approximated by the metric 
 \beq\label{rindlop}
 ds_{\rm op}^2 \approx -dt^2 + dz^2 + e^{4\pi T (z-z_\beta)} d\ell^2\;, \qquad z>z_\beta \;,
 \eeq
with the horizon sitting at $z=\infty$.  Notice that the spatial sections of this metric are, locally in $\ell$-space, $(d+1)$-dimensional hyperboloids with radius $\beta/2\pi$, i.e. under the change of variables $y=(2\pi T)^{-1} e^{-2\pi T(z-z_\beta)}$ we have, on $\ell$-distance scales where the  QFT metric (\ref{qftm}) is approximately flat, 
\beq\label{ophy}
ds_{\rm op}^2 \approx  -dt^2 + \left({\beta \over 2\pi}\right)^2 \,ds^2_{{\bf H}^{d+1}}\;,
\eeq
where 
\beq\label{hy}
ds^2_{{\bf H}^{d+1}} = {dy^2 + d\ell^2 \over y^2 }
\eeq
is the metric of the unit radius hyperbolic space in $d+1$ dimensions. In these coordinates, the near-horizon region is defined by $2\pi y\leq \beta$, with the horizon itself sitting at $y=0$.  Therefore, trajectories of photons in the near-horizon optical metric  trace a geodesic flow on the hyperbolic space ${\bf H}^{d+1}$. 

 The optical depth of the Rindler region acquires the simple interpretation of an `optical distance' to the stretched horizon: 
\beq\label{reftt}
\tau_* \sim z_* - z_\beta \sim \beta \,\log \left({\rho_\beta - \rho_0 \over \rho_* - \rho_0}\right) \sim  -\beta\,\log (h'_0\, \ell_{\rm P})\sim \beta\,\log\left( {\sqrt{F_0 }\over T \,\ell_{\rm P}}\right)\;,
\eeq
where we have neglected numerical factors of $\CO(1)$.

In order to translate (\ref{reftt}) into dual QFT variables, we compute the entropy,  or   horizon volume in Planck units:
$$
S\sim {1\over \ell_{\rm P}^d} \int d^d \ell  \left(\sqrt{F_0}\right)^d \sim V \left({\sqrt{F_0} \over \ell_{\rm P}}\right)^d\;,
$$
with $V = \int d^d \ell$ the volume  in the QFT metric. With these conventions, the entropy in a QFT volume of one thermal length, 
$V_{\rm cell} \sim \beta^{d}$, is given by $S_{\rm cell} = S /n_{\rm cell} \sim (T \ell_{\rm P} / \sqrt{F_0})^{-d}$, where
$n_{\rm cell} = VT^d$ is the number of `thermal cells'. Returning now to (\ref{reftt}), we find,  up to $\CO(1)$ numerical coefficients, the following expression in dual QFT variables
\beq\label{rru}
\tau_* \sim \beta \,\log\left({S \over n_{\rm cell}}\right)= \beta \,\log \,(S_{\rm cell})\;,
\eeq
as claimed in the Introduction. We find that   the no-cloning time scale $
\tau_*$  is sensitive to the parameters of a single thermal cell. This is actually natural, since the emission of a typical  Hawking bit can be  localized spatially on the horizon within a region of size $\beta$ in $\ell$-space. Therefore, the behavior of regions larger than a thermal cell  have a less direct relevance to the issue of  information retrieval by a {\it single} Hawking bit.
 
 The entropy per thermal cell, $S_{\rm cell}$,  estimates  the number of  microscopic  QFT degrees of freedom participating on a thermal state at temperature $T$, and supported on length scales of order $\beta$, so that we may use the notation $S_{\rm cell} = N_{\rm eff} (T)$. For  a conformal field theory $N_{\rm eff}$ is asymptotically independent of the temperature and proportional to the central charge, of order $N_{\rm eff} \sim N^2$ for CFTs based on Yang--Mills theories. For non-conformal theories it gives a sort of `running' central charge with non-trivial scale dependence. This  running central charge also controls other observables related to counting of degrees of freedom, such as the entanglement entropy computed in the gravity prescription \cite{ryu}, (see  \cite{fuertes} for such a discussion). 

\section{ Scrambling And Hard Chaos}

\noindent

The heuristic arguments leading to the $\tau_* = \beta \,\log(S_{\rm cell})$ time scale, essentially causality bounds, 
are based on a phenomenological treatment of the horizon. Standard large-$N$ formulations of the AdS/CFT rules are not sensitive to  the finiteness of a horizon entropy, implying infinite values for the  scrambling time scale, as well as the (Heisenberg) time scale of Poincar\'e recurrences. The only horizon time scale that is visible at $N=\infty$  is the quasinormal behavior of local correlation functions, i.e. the exponential decay at large times
\beq\label{qnb}
\bra \CO(t) \CO(0)\ket \sim e^{- t/\tau_\beta} \;, \qquad t\gg \tau_\beta \sim \beta\;. 
\eeq
This behavior is dual to the familiar no-hair property of bulk horizons and characterizes the loss of information in the $N=\infty$ system.\footnote{The quasinormal time scale $\tau_\beta \sim \beta$ also controls the large-$N$ time-dependence of certain non-local (but sharp) operators, such as the entanglement entropy (cf. \cite{espe}).}  In terms of the bulk description, the quasinormal behavior is associated to the continuous spectrum of bulk fields in the vicinity of the horizon, in particular, the single-particle frequency spectrum $\omega$
is obtained from the eigenvalues of the operator (cf. \cite{brickwall, horyo})
\beq\label{eigenv}
{\hat \omega}^2 =-{d^2 \over dz^2} + V_{\rm eff} (z)\;,
\eeq
where the effective potential has a universal form near the horizon of (\ref{holob}), 
 \beq\label{rindl}
V_{\rm eff}(z)_{\rm Rindler} \propto \left(F_0 \,m^2 + {c\over \varepsilon^2}\right) \,\exp\left(-{4\pi T z}\right)\;, \qquad z\gg z_\beta \sim \beta\;,
\eeq
featuring an exponential barrier controlled by the temperature $T$. Here $m$ is the mass of the bulk field and 
$p_\ell^2 \sim c/\varepsilon^2$ is the effective momentum-squared introduced by  the UV localization of the particle probe within a cutoff length $\varepsilon$, as  measured by the $\ell$ coordinate. Local CFT correlation functions are defined as boundary correlation functions of such bulk fields, and the quasinormal behavior (\ref{qnb}) is related to tunneling through the exponential barrier seen in (\ref{rindl}). 

The dynamical time scales at the horizon are rendered finite by specifying a phenomenological `stretching', whose crudest version is the brick-wall model \cite{brickwall}, consisting on a sharp mirror at Planckian distance from the horizon. This simple boundary condition discretizes the spectrum of fields in the right amount to account for the finiteness of the entropy, even for a  thermal state of free quantum fields. Such a finite entropy   turns out to be {\it extensive} in the optical metric, 
$$S\sim V_{\rm op} \;T^{\,d+1}\;,
$$
 with the optical volume of the region outside the stretched horizon scaling as $V_{\rm op} \sim \beta V N_{\rm eff}$ (cf. \cite{optical, horyo}). This simple argument shows that the optical metric provides a good geometrical intuition for the `phase space size' of the thermal atmosphere, i.e. the number of states supported by the system. In the conformally related frame of the optical metric, the Hartle--Hawking state of quantum fields is a homogeneous thermal state of constant temperature $T$ over a cut-off version of the  hyperboloid (\ref{hy}).

This very crude model reproduces the right time scale of recurrences \cite{russ}. It also introduces a ballistic free-fall time scale of order $\tau_*$, which is not obviously related to `scrambling' since interactions of localized probes with the thermal atmosphere are suppressed by powers of $1/N_{\rm eff}$. 
On the other hand, low-energy effective field theory must be violated at the stretched horizon, so that the $1/N_{\rm eff}$ estimate for the interactions breaks down at the wall. This is one of the basic tenets of black-hole complementarity \cite{bhcompl}, and can be seen in this context by noticing that a Planck-mass field $m\sim m_{\rm P}$ does make a contribution of order $T^2$ to the effective potential (\ref{rindl}) when dragged to the stretched horizon $z\sim z_*$. Hence, we can scramble the probe by letting it interact strongly at the wall, after a ballistic glide of $\tau_*$ duration.  

A naive treatment of such a `wall-scrambling' seems to suggest a behavior of `slow-scrambler' type. In the optical-frame description, the state of the bulk quantum fields is thermal with uniform temperature $T$. Once the probe reaches the stretched horizon, if confined there, one can see it scatter on time scales of order $\beta$, exercising a random walk through the stretched horizon with step $\beta$. Since the QFT thermal cell has optical size $\beta (N_{\rm eff})^{1/d}$, the time to scramble the whole QFT thermal cell is of order 
$\beta (N_{\rm eff})^{2/d} \sim \beta (S_{\rm cell})^{2/d}$, i.e. the characteristic behavior of a slow scrambler.\footnote{ The same result follows by thinking in terms of local static observers at the stretched horizon, who measure local Planck temperatures. In this frame, the Planck-size probe will execute a random walk of Planck step. Since there are $S_{\rm cell}$ Planckian cells in a single QFT thermal cell, the random walk will cover it in a (local) time 
  $\ellp (S_{\rm cell})^{2/d}$. Upon redshifting back to QFT-time, $t$, we see that a QFT thermal cell is scrambled in a time of order $\beta (S_{\rm cell})^{2/d}$.} 

In the next section we argue that this estimate is not quite correct. The reason is that, while the collisions are happening at the stretched horizon, the ballistic glide between collisions is not necessarily confined to the stretched horizon hypersurface. In  fact,  as already pointed out in section 2, the faster free-fall motion between two points on the stretched horizon takes a path {\it above} the stretched horizon. This fact, combined with the peculiar geometry of the Rindler space, will suffice to provide a kinetic model of the thermal atmosphere with fast `scrambling' features.

\subsection{Optical Metric And The Chaotic Billiard Model}

\noindent

Let us consider a Planck-localized probe following null geodesics in the Rindler region, punctuated by collisions of $\CO(1)$ strength at the stretched horizon (we neglect the collisions of $\CO(1/N_{\rm eff}) $ strength within the bulk of the thermal atmosphere). In order to benefit from the geometrical intuition, we notice that any such piecewise-null trajectory defines a corresponding piecewise-null trajectory on the conformally-related optical manifold (\ref{rindlop}) with hyperbolic spatial sections. Therefore, we can use the optical metric to discuss the dynamical 
time scales of such an ideal probe motion. The advantage of  this  description is the uniform proportionality between time scales for null propagation and `optical' length scales. 

The optical box supports a thermal state of bulk fields at {\it uniform} temperature $T$, which can be approximated as an ideal gas  (up to small $1/N_{\rm eff}$ effects,) except at the stretched horizon, where the thermal fields are regarded as strongly interacting, with a bulk thermal cell of (optical) size $\beta$. 

Consider a localized probe sent radially inwards with a maximal (Planck) bulk resolution at the top of the Rindler region, as measured in the physical metric (\ref{holob}), i.e. the initial resolution $\delta \ell_i$  in the $\ell$ coordinate satisfies
$$
F_0 \,(\delta\ell_i)^2 \sim \ellp^{\,2}\;.
$$
The geometrical stretching of the optical metric implies that such a probe is smeared over a region whose optical size in the $\ell$ directions is of order
 $$
\delta s_{\rm op} \sim { \delta \ell_i \over \sqrt{h_*}} \sim {\ellp\over \sqrt{F_0 h_*}}  \sim \beta
 $$
  on arrival to the stretched horizon. The thermal state of strongly interacting degrees of freedom at the stretched horizon has thermal length of $\CO(\beta)$ (again, in the optical metric), which determines the effective interaction length of probe scattering at the stretched horizon. Since   the effective resolution in the impact parameter for these scattering events is also of $\CO(\beta)$,  we conclude that the outgoing scattering angle at each collision has maximal uncertainty.

  In between scattering events the probe is assumed to glide freely on the Rindler region. Maximizing the glide velocity, we can assimilate the motion to a random walk on a $(d+1)$-dimensional hyperboloid with steps made of geodesic arcs on the  spatial sections of the optical geometry:
\beq\label{hyps}
ds^2_{\rm op} \approx -dt^2 + \left({\beta \over 2\pi}\right)^2 {dy^2 + d\ell^2 \over y^2}\;.
\eeq
 
In the Poincar\'e coordinates of (\ref{hyps}), each free glide between collisions is a circular arc with radius $\Delta y \sim \Delta \ell /2$ if the collision points are separated a distance $\Delta \ell$ in the QFT metric (cf. Figure 2). 
The maximum extent of a circular glide is given by  arcs of radius $y_{\rm max} \sim \beta$. The reason is that the scarce\footnote{The proportion of solid angle enclosing those trajectories which are `vertical' enough to  hit the end of the Rindler region is of the order of $\delta^d / {\rm Vol}({\bf S}^d) \sim (h_*)^{d/2} \sim 1/N_{\rm eff}$.} trajectories with 
larger radii hit the end of the Rindler region and are effectively reflected back by the confining wall of the asymptotic AdS region, the resulting complete trajectory returning to  the stretched horizon at $\Delta \ell < \beta$. We thus conclude that, after a few scattering events, there is complete uncertainty about the position of the probe within a distance
scale of the order of the QFT thermal length. Since each maximal  glide through the optical hyperboloid takes a time of $\CO(\tau_*)$, we conclude that the `position' scrambling of the Planckian probe within  the thermal cell has been achieved in times of order $\tau_* \sim \beta \log(S_{\rm cell})$.

\begin{figure}
  \label{billar}
  \begin{center}
    \epsfig{file=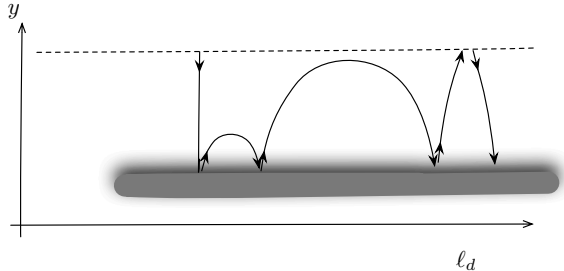, width= 9.5cm}
    \caption{\small Near-horizon random walk of a localized probe by scattering at the stretched horizon, pictured in the Poincar\'e coordinates of the spatial optical metric (\ref{hyps}). Free paths between successive collisions are circular arcs, with maximal radius $\Delta y_{\rm max} \sim \beta$, since longer glides are reflected back by  the asymptotic AdS potential  well which starts at the edge of the Rindler region, represented in this picture by a dashed line. }
  \end{center}
\end{figure}

The kinetic model presented  here regards the near-horizon region as a chaotic billiard over a negative-curvature space, one of the early  examples of hard chaos in classical mechanics \cite{Gutzwiller, Voros}. Indeed, if we substitute the stretched horizon by
a random arrangement of hard reflecting spheres of size $\beta$ in the optical metric, we get essentially the same picture. The Lyapunov exponent, defined locally in terms of the divergence rate of geodesics, is of order $\lambda_{\rm L} \sim \beta^{-1}$. The Lyapunov time, defined as the time for complete erasure of localization information, from an initial resolution $\delta\ell_i$ to a final resolution of one thermal length $\delta\ell_f \sim \beta$,  is of order
$$
\tau_{\rm L} \sim \lambda_{\rm L}^{-1} \log(\delta\ell_i /\delta\ell_f)\sim \beta\,\log(\ellp/\beta\sqrt{F_0}) \sim \tau_*\;,
$$
 since
complete ignorance over the position of the Planck-size probe on the extent of a thermal cell is achieved after $\CO(1)$ collisions, with characteristic collision time of $\CO(\tau_*)$. 

Our proposed  `ballistic catalysis'  of the scrambling is only effective as long as the glides are large geodesic arcs on the
hyperbolic optical geometry (\ref{hyps}). Since the finite extent of the Rindler region imposes an effective cutoff to the size of these
arcs, on time scales larger than $\tau_*$ the system behaves as a slow scrambler with effective time step $\tau_*$. Therefore, the whole system with $n_{\rm cell}$ thermal cells is scrambled in a time $\tau_s \sim \tau_* (n_{\rm cell})^{2/d}$. In this way we arrive at our main result of the paper:  the scrambling time of a general horizon as a function of the entropy and the effective number of degrees of freedom 
 \beq\label{fastsc}
\tau_s \sim \beta\,\log(N_{\rm eff}) \, (n_{\rm cell})^{2/d} \sim \beta\,\log(N_{\rm eff}) \,\left({S \over N_{\rm eff}}\right)^{2/d}\;.
\eeq

 \section{Conclusions}
 
 \noindent
 
 We have discussed aspects of the fast-scrambling time scale, introduced in \cite{Hayden, Sekino} as proportional to the logarithm of the entropy, from the viewpoint of the AdS/CFT correspondence. 
 
 We first pointed out that a naive implementation of the no-cloning bound singles out the free-fall time scale across the Rindler region, $\tau_* \sim \beta\,\log(S_{\rm cell})$, as the important figure of merit. We have shown that there is a natural geometrical interpretation of this quantity in terms of the `depth' of the stretched horizon in the optical metric for an arbitrary AdS/CFT background.  
 
 The dependence of $\tau_*$  on the entropy of a single thermal cell, $S_{\rm cell}$, deserves special comment. In particular, this result was derived under the assumption that the QFT thermal state is extensive on length scales larger than the inverse temperature. For very cold states it may happen that the system is actually smaller than its thermal length. One interesting example is given by the Reissner--Nordstrom black hole, for which the entropy remains finite in the limit of divergent thermal length: $\beta \rightarrow \infty$. Computing the optical depth for such a near-extremal black hole one finds $\tau_* \sim \beta\,\log(S)$, rather than (\ref{univ}) (notice that both laws are equivalent for Schwarzschild black  holes, since those have a single thermal cell on their horizon.) It turns out that the rules of the AdS/CFT correspondence ensure that it is the total entropy, rather than the entropy per thermal cell,  the relevant quantity when the scrambler is smaller than its thermal length \cite{usfut}. 
 
 The hyperbolic character of the near-horizon optical metric suggests that fast scrambling {\it may} be compatible with a purely kinetic picture of information diffusion. In particular, local probes scattering at the stretched horizon with null free paths do scramble within a single thermal cell in just a few collision times, because the negative curvature amplifies any initial resolution error of the probe, making the near-horizon region into an effective {\it chaotic billiard} with Lyapunov time of order $\tau_*$. On scales larger than a single thermal cell, the subsequent scrambling follows the pattern of standard slow scramblers with time step of $\CO(\tau_*)$, leading to our main estimate 
 \beq\label{maine}
 \tau_s \sim \tau_* \,(n_{\rm cell})^{2/d}
 \eeq
 of the scrambling time of a general fast scrambler with $n_{\rm cell} \geq 1$. 
 
 Our result indicates that classical chaos may be a key element in making compatible the built-in large-$N$ locality of bulk dynamics with the apparently gross violation of locality implied by the very fast scrambling of information at {\it generic} horizons. This includes very interesting physical situations of a different kind, such as de Sitter spacetime, for which there is no concrete holographic description. 
 
 On the other hand, it is not obvious that the kinetic notion of scrambling adopted here is the relevant one from the point of view of the issues of black-hole information retrieval, as  laid down in \cite{Hayden}. In particular, while classical chaos implies a certain notion of scrambling in position space for local probes, over times of order $\tau_*$ only a few scattering events can take place and thus only a small subspace of the complete Hilbert space has a chance to entangle with the probe.  In addition, non-local probes such as extended branes may behave in quite a different way from the `billiard paradigm' outlined here, so that strongly nonlocal physics on the scale of the thermal length may still be required to achieve scrambling in the sense of the quantum entanglement test \cite{Hayden}. 
 
It is conceivable that the details of the scrambling will depend on the structure of the state, such as the initial degree of localization of the information. In this context, the mechanism proposed here is mainly a fast way of homogenizing the quantum state over the thermal cell,  which may be subsequently scrambled non-locally  throughout the whole Hilbert space of dimension $\exp(N_{\rm eff})$, as in a matrix model. Both processes could be accomplished in times of $\CO(\tau_*)$, so that the complete scrambling time  would still satisfy  $\tau_s \sim \tau_*$ within numerical factors.

 We leave these and other fascinating issues for future work. 
 It would be of great interest  to independently check (\ref{fastsc}) in a toy model of the type studied in 
 \cite{festuccia}, although the relevance of the Rindler region in the heuristics behind (\ref{fastsc}) suggests that CFTs with large gaps in the conformal dimensions (cf. \cite{polc}) are needed to see the $\log(N_{\rm eff})$ term. 
More generally, we can only expect that quantum chaos will enter as a key idea in the characterization of black-hole information processing. 

 \vspace{0.5cm}
 
{\bf Acknowledgements:}  We are indebted to C. G\'omez, E. L\'opez and E. Rabinovici for useful discussions. 
 The work of J.L.F.B. was partially supported by MEC and FEDER under a grant FPA2009-07908, the Spanish
Consolider-Ingenio 2010 Programme CPAN (CSD2007-00042) and  Comunidad Aut\'onoma de Madrid under grant HEPHACOS S2009/ESP-1473. J.M.M. is supported by a FPU fellowship from MICINN.

\end{document}